\documentclass[11pt]{article}
\usepackage{hyperref}
\pdfoutput=1
\begin{document}
\title{Falling Flexible Sheets}
\author{Silas Alben \\
\\\vspace{6pt} School of Mathematics \\ Georgia Institute of
Technology, Atlanta, GA 30332-0160, USA}
\maketitle
%% The abstract (in this file, and that submitted as text to arXiv) should
%include the exact phrase
%% "fluid dynamics video" or "fluid dynamics videos"
\begin{abstract}
We present a fluid dynamics video showing simulations of flexible bodies falling in an inviscid fluid. Vortex sheets are shed from the trailing edges of the bodies
according to the Kutta condition. The basic behavior is a repeated series
of accelerations to a critical speed at which the sheet buckles, and rapidly
decelerates, shedding large vortices. Examples of persistent circling,
quasi-periodic flapping, and more complex trajectories are shown.
\end{abstract}
% main text

The video is shown in
\href{http://ecommons.library.cornell.edu/bitstream/1813/17067/2/BuckleFlapMpeg2.mpg}{high-resolution} and
\href{http://ecommons.library.cornell.edu/bitstream/1813/17067/3/BuckleFlapMpeg1.mpg}{low-resolution} files.

The first clip is entitled ``{\bf Falling flexible sheet
trajectories: Example of buckling while falling}.'' The moving solid orange line
is a flexible fiber, falling under gravity, and shedding a vortex sheet (blue line) from its
trailing edge. The two control parameters are the sheet density normalized by fluid
density ($R_1$, 0.3 here) and the sheet rigidity normalized by fluid inertia ($R_2$, 2.4 here).
The basic behavior is a repeated series of accelerations to a critical speed at which the sheet buckles, and rapidly decelerates, shedding large vortices. The still frames which
surround the moving picture give sample trajectories for many different initial falling angles
and different sheet
rigidities. These paths show a diversity of punctuated falling and circling behaviors (circling is seen for
$R_2$ equal to 10 and above). The still frame labeled ``$R_1=0.3$, $R_2=2.4$'' shows in light blue
the trajectory traced by the orange fiber as it falls.

The second clip, ``{\bf Falling flexible sheet
trajectories: Examples of quasi-periodic flapping},'' shows an alternative falling behavior.
For a range of smaller $R_1$ and $R_2$ (two examples are shown), the body flaps steadily as
it falls. The example to the left is an asymmetric flapping state. The example to the right
shows symmetric flapping with a simple period. The drag encountered
by these flapping bodies balances the acceleration from gravity. The still panels again
show examples of different falling trajectories as parameters are varied. The blue trajectories
in the still frame labeled ``$R_1=0.3$, $R_2=1$''correspond to states of flapping while falling,
encountered for many different initial falling angles.

\end{document}